\documentclass[5pt]{aastex61}
\renewcommand{\baselinestretch}{1.5}
\usepackage[utf8]{inputenc}
\usepackage{nameref}
\usepackage{wasysym}
\usepackage{makecell}
\usepackage{sidecap}
\usepackage{soul}
\usepackage{comment}
\usepackage[title]{appendix}

\usepackage{fancyvrb}
\usepackage{BibDef}

\graphicspath{{img/}}

\begin{document}
\title{Giant planet formation models with a self-consistent treatment of the heavy elements}

\author{Claudio Valletta and Ravit Helled}
\affiliation{Center for Theoretical Astrophysics \& Cosmology, 
Institute for Computational Science, 
University of Zurich, Zurich, Switzerland}

%\author[]{Claudio Valletta}
%\affiliation{Center for Theoretical Astrophysics $\&$ Cosmology, Institute for Computational Science, University of Zurich, Switzerland}

%\author[0000-0001-5555-2652]{Ravit Helled}
% \affiliation{Center for Theoretical Astrophysics $\&$ Cosmology, Institute for Computational Science, University of Zurich, Switzerland}

% List of institutions
%$^{1}$Center for Theoretical Astrophysics and Cosmology, Institute of Computational Science, University of Zurich,
%Winterthurerstrasse 190, 8057 Zurich, Switzerland\\
%}
\accepted{for publication in ApJ}
\begin{abstract}

We present a new numerical framework to model the formation and evolution of giant planets. The code is based on the further development of the stellar evolution toolkit Modules for Experiments in Stellar Astrophysics (MESA).
The model includes the dissolution of the accreted  planetesimals/pebbles, which are assumed to be made of water, in the planetary gaseous envelope, and  the effect of envelope enrichment on the planetary growth and internal structure is computed self-consistently. 
We apply our simulations to Jupiter and investigate the impact of different heavy-element and gas accretion rates on its formation history. 
We show that the assumed runaway gas accretion rate significantly affect the planetary radius and luminosity.
It is confirmed that heavy-element enrichment leads to shorter formation timescales due to more efficient gas accretion. 
We find that with heavy-element enrichment Jupiter's formation timescale is compatible with typical disks' lifetimes even when assuming a low heavy-element accretion rate (oligarchic regime). 
Finally, we provide an approximation for the heavy-element profile in the innermost part of the planet, providing a link between the  internal structure and the planetary growth history.

\end{abstract}

\keywords{methods: numerical --- planets and satellites: formation, gaseous planets --- protoplanetary disks -- planet-disk interactions}

\section{Introduction} \label{sec:introduction}

Understanding the origin of giant planets is a key goal in planetary science.  
In particular, it is of great interest to link formation models with the planetary composition and internal structure. It is therefore required that giant formation models would provide predictions on the expected planetary bulk composition and the distribution of the heavy elements in the deep interior.

In the core accretion model \citep{pollack1996}, the leading scenario for giant planet formation \citep{Helled14}, the growth of a giant planet begins with the formation of a heavy-element core.
Once the core reaches about Mars' mass, its gravity is strong enough to accrete hydrogen and helium (hereafter H-He) gas from the proto-planetary disk. 
The proto-planet then continues to grow by accreting both solids (heavy elements), in the form of planetesimals (0.1 - 100 km sized objects), and H-He until the so-called crossover mass is reached \citep{lissauer2009,Ginzburg2019}  and runaway gas accretion takes place. 

An alternative scenario to core formation via planetesimal accretion is pebble accretion. Pebbles are small solid particles with typical sizes of 1-10 cm  \citep{Ormel10,Lambrechts12,Lambrechts14,Bitsch15}. Due to their small sizes, pebbles experience significant gas drag that slows them down. This leads to a flux of pebbles that can be accreted very efficiently by the growing planet. The pebble accretion rate is high and is estimated to be $\sim 10^{-5}$ M$_\oplus$/yr \citep{morbidelli15}. Pebble accretion is stopped by the perturbation induced by the planet on the disk. This occurs when the proto-planet reaches the so-called pebble isolation mass. At Jupiter's location of 5.2 AU, the pebble isolation mass is $\sim$ 20 M$_\oplus$ \citep{Lambrechts14}.

Independently on the size of the accreted solids, a common assumption in giant planet formation models is that the heavy elements reach the planetary center ("core").  
%while H-He stays in the envelope.
This, however, is a strong simplification because the accreted solids can undergo ablation and fragmentation, and therefore much of their mass (if not all) is deposited in the envelope. Indeed, as shown in previous studies \citep{pollack86,Brouwers2017,Alibert17,Lozovsky17} when the core mass is between $1 - 3$ M$_\oplus$ solids are expected to dissolve in the gaseous envelope. 
%The dissolution of solids plays a key role in the early formation stages, although heavy-element accretion can  also occur during runaway gas accretion  \citep{Shibata19}.
Therefore, only for the naive scenario in which all the accreted heavy elements are assumed to go to the center the heavy-element mass in the planet is comparable to the core's mass.
%Therefore, only for the naive scenario in which  heavy elements do not fragment and/or \textbf{vaporises} the total heavy-element mass in the planet is comparable to the mass of the central core. 
This is clearly not a realistic scenario as shown by various previous studies \citep{HelledStevenson17, Brouwers20,Brouwers18,VenturiniHelled17,Bodenheimer18}
\par

Previous studies including heavy-element enrichment were mainly focused on analysing the ablation of planetesimals by the proto-planet envelope, inferring a maximum core mass of giant planets, or computing the planetary evolution with the presence of a heavy-element gradient. 
In fact, already in \cite{pollack86} where planetesimal accretion was considered, a maximum core mass between $1$ and $3$ M$_\oplus$ for $100$ km-sized planetesimals was derived. 
\cite{Lozovsky17} investigated the heavy-element distribution in proto-Jupiter accounting for different solid surface densities and planetesimal sizes. A maximum core mass of 2-3 M$_\oplus$ was found, with the rest of the heavy elements having a gradual distribution in the planetary deep interior.
\cite{Venturini16} presented a calculation that includes the heavy-element enrichment in the planetary growth. 
It was assumed that the proto-planet accretes planetesimals composed of rock and water and that the rocks sink to the center (i.e., joining the core) while the water is dissolved in the envelope. The water in the envelope was assumed to be instantaneously uniformly mixed, so the envelope composition is always homogeneous but enriched in water. It was shown that the planetary formation timescale can be  significantly reduced when envelope enrichment is considered. 
The same computational method was used in \cite{VenturiniHelled17}  where the aim is to predict the occurrence rate of mini-Neptunes, which is found to be larger if heavy-element enrichment is considered. 
%In \cite{Vazan17} was presented the evolution of Jupiter assuming a primordial structure with a total heavy elements mass of $40$ M$_\oplus$ distributed non uniformly through the planet's radius. it was shown that mixing is expected to occur in the outer part, leading to an adiabatic outer envelope, while the composition gradient in the deep interior that persists.
Finally, \cite{Bodenheimer18} analyzed the formation of the Kepler 36 system, modeling the dissolution and fragmentation of 100 km rocky planetesimals. They included mass loss induced by stellar XUV radiation as well as planetary migration. 
This simulation could reproduce the mass and radius of Kepler 36-c. It was found that, when heavy-element enrichment is included, in the region of  the envelope which is highly enriched with heavy elements  (outer core in their paper) the temperature increases significantly, while the density decreases.
%the temperature increases significantly in the region where the majority of the heavy elements are present, while the density decreases. 
The inferred lower densities imply that only about half of the H-He mass typically required is needed to fit the present-day mass and radius measurements. 
\par

Until now, heavy-element enrichment has not been implemented in a self-consistent way in planet formation simulations.  
This is due to the numerical challenges linked to this problem: at each time-step the deposited heavy-element (high-Z material) mass and energy must be computed. The presence of heavy elements in the envelope then has to be included  in the opacity and Equation of State (EOS) calculation in order to compute the envelope's structure correctly. 
In our previous work \citep{Valletta2019} (hereafter VH19) we performed a first step towards a self-consistent model for giant planet formation where we analyzed planetesimals' mass loss with pre-computed structure models of the planet at different times. 
We investigated how ablation and fragmentation depend on various parameters such as the planetesimal's size and composition, different ablation efficiencies ($C_h$ factor)  as well as the impact of assuming different fragmentation models.  
 
In this work we present a giant planet formation model with a self-consistent treatment for the heavy elements deposition. The heavy elements are represented by water (i.e., H$_2$O).
The paper is organized as follows:
In section \ref{sec:methods} we present the numerical methods, and provide detail description of the modifications implemented in MESA. 
%At each time-step, we compute the heavy-element mass and energy deposition, which leads to a non-uniform  enrichment of the envelope. We then include the presence of high-Z material in the envelope's EOS and opacity calculations that govern the evolution of the planet. 
In section \ref{sec:solidandgas} we investigate the effect of different heavy-element (solid) and gas (H-He) accretion rates without including heavy-element enrichment. 
We show that different runaway gas accretion rates strongly affect the planetary radius and luminosity.
In section \ref{sec:planetesimals}, \ref{Sec-oligarchic} and section \ref{sec:pebbles} we implement heavy-element enrichment and  investigate its effect on the Jupiter's growth considering both planetesimal and pebble accretion. We also show the impact of using different fragmentation models. 
We investigate the effect of heavy-element  enrichment on the planetary formation time-scale and the inferred primordial internal structure assuming both pebble and planetesimal accretion.
Finally, in section \ref{sec:planetarymemory}  we provide an approximation for the heavy-element profile in the planet that links the internal structure with the planetary growth history. 
Our conclusions are summarized in section \ref{conclusion}.

\clearpage
\section{Methods}
\label{sec:methods}

We employ the MESA toolkit (\citealt{Paxton2010},\citealt{Paxton2013},\citealt{Paxton_2018}) (version 10108) to evolve a proto-planet embedded in a protoplanetary disk. The simulations begin with a heavy-element core with a mass of $M_{icore} = 0.01 M_{\oplus}$ (see \citealt{Chen2016} or \cite{malsky2020} for a discussion on low-mass MESA initial model) and a negligible envelope mass.
The planet is assumed to be in hydrostatic equilibrium and spherically symmetric. 
The original equation of state (EOS) present in MESA is not applicable to describe a mixture of water and H-He at a temperature-pressure regime that is  typical of giant planets. We therefore use the EOS module developed by \cite{Muller20} to properly model a mixture of water and H-He in planetary conditions. Details on the implementation of the high-metallicity EOS in MESA can be found in \cite{Muller20} and references therein. 
Although it is clear that the heavy elements could consist of a mixture of elements, for simplicity in this work the accreted planetesimals and pebbles are assumed to be composed of pure water (H$_2$O). 
Large fractions of water are justified by the assumed formation locations of 5.2 and 10 AU which are clearly beyond the water ice line (e.g., \citealt{Sasselov2000}). 
Representing heavy elements with water is a common assumption in giant planet formation and evolution simulations \citep{Vazan13,Vazan15,Vazan16,Venturini16,Yann18}. This is because the equation of state of water has been widely investigated and the available tables cover a large-enough range of pressures and temperatures expected in planetary interiors  \citep{More88,French09,Mazevet19}. 
Since in this study the inclusion of heavy elements in the model corresponds to the modification of the EOS, as long as the heavy elements are represented by materials that are significantly heavier than H-He the inferred results do not change significantly \citep{Vazan13, Muller20}. 
We hope to consider various heavier elements in future research.

We use the standard low temperature \citep{Freedman2014} Rosseland tables for the gas opacities by using the options in the
\verb!inlist_evolute! file \verb!kappa_lowT_prefix = lowT_Freedman11! \verb!kappa_file_prefix = gs98!. 
The gas opacity is scaled with metallicity, where  higher heavy-element 
fractions typically increase the opacity. Therefore, when we 
deposit heavy-element mass in a layer, the gas opacity is changed 
accordingly.
Determining the grain opacity is challenging since it 
depends on many parameters, such as the number of grains present in a 
layer, their composition, shape, and size distribution. 
%and shape, which depends on the dust microphysics. 
The grain microphysics is also expected to change with time due to physical processes such as coagulation, sedimentation, evaporation, etc. Therefore for simplicity, the dust grain opacity is simply calculated using the prescription of \citet{Valencia2013} where the grain opacity is not calculated  directly, but is extrapolated  from the opacity tables of \citet{Alexander94}. We insert this fit into the 
\verb!other_opacity! subroutine.
%The dust opacity derived by \citealt{Valencia2013} does not explicitly  scale with metallicity.
%We include the effect of dust grains to the opacity following \citealt{Valencia2013} by inserting their fit to the \cite{Alexander94} dust opacity into the \mintinline{fortran}{other_opacity} subroutine. 
The outer temperature and pressure are set to the disk's pressure and temperature \citep{Chiang10} at the planet's location (\verb!fixed_T_and_P!).

The heavy-element compact core is modeled using an inner boundary condition which sets the mass, radius, and luminosity at the base of the envelope. We use the tables from \citealt{Chen2016} to determine the core's radius at a given mass. The infalling planetesimals typically reach the core only at the very early stages of the formation process when the envelope's mass is negligible \citep{Valletta2019, Brouwers20}.

%The inner boundary condition is provided by the small heavy-element core, which is growing via heavy-element accretion at the very beginning of the simulation when the envelope's mass is negligible and the accreted solids are not affected much by the gas and therefore are added to the core.  
%The core is modeled according to \citealt{Chen2016}, where the core's radius is determined by its mass, the temperature and pressure of the innermost layer of the envelope. 

To simulate giant planet formation with heavy-element  enrichment, we implemented three main sub-routines into MESA: (1) a sub-routine that computes the heavy-element accretion rate; (2) a sub-routine that computes the interaction of the accreted solids with the envelope, including mass ablation and fragmentation; and (3) a sub-routine that controls the gas accretion rate. 

The results of each of the three sub-routines affect the others: the heavy-element accretion rate drives the accretion of gas, which in return affects the ablation.  On the other hand, the mass deposition of heavy elements determines the envelope's composition which affects both the gas and solid accretion rate. 
Finally, the total envelope's mass (which depends on the gas accretion) determines the feeding zone of the proto-planet, and therefore the heavy-element accretion rate.  
The sub-routines are called at each time-step by using the \verb!other_wind! sub-routine in MESA. 

%We now describe in detail the implementation of each of these subroutines.

%\subsection{Solid accretion rate}
%\label{sec:solidaccretionrate}

\subsection{Heavy-element accretion rate}

We implemented different solid (heavy-element) accretion rates: (1) the planetesimal accretion rate from \cite{pollack1996}, (2) the oligarchic accretion rate based on the N-body simulations from  \cite{Fortier07}; (3) the accretion rate of  \cite{Shiraishi08} which corresponds to a scenario with the absence of \textit{phase-2}, and (4) the pebble accretion rate of \cite{Lambrechts14} where we include the 3d effects as shown by \cite{morbidelli15} (Equation~17). Table~\ref{table1} lists the heavy-element accretion rates used in this study. 
The planet's capture radius is determined using the prescription of \cite{Ikoma08} at the beginning of each time-step.

\begin{table}[h!]
\begin{center}
 \begin{tabular}{c c c c c } 
 \hline
 Heavy-element accretion rate & Reference & Size & Density & Disk's viscosity  \\ 
 \hline\hline
 $\dot{M}_{Z}$1 & \cite{pollack1996}, their Equation~1 & 100 km &$\sigma =$ 10 g cm$^{-2}$ & / \\ 
 $\dot{M}_{Z}$2 & \cite{Lambrechts14}, their Equation~31 & $\sim$ 1 cm& $\Sigma =$ 50 g cm$^{-3}$& $\alpha = 10^{-3}$ \\
  $\dot{M}_{Z}$3 & \cite{Lambrechts14}, their Equation~31 & $\sim$ 1 cm&$\Sigma =$ 50 g cm$^{-3}$& $\alpha = 10^{-5}$\\
 $\dot{M}_{Z}$4 & \cite{Fortier07}, their Equation~10 & 100 km & $\sigma =$ 10 g cm$^{-2}$ & /\\
 $\dot{M}_{Z}$5 & \cite{Shiraishi08}, their Equation~24 and 25 & 100 km & $\sigma =$ 10 g cm$^{-2}$ & /\\
 \hline
\end{tabular}
\end{center}
\caption{The various heavy-element accretion rates used in this work.}
\label{table1}
\end{table}

When the value of the heavy-element accretion rate is determined, we accordingly update the inner boundary condition for the structure equations: 
\begin{verbatim}
s%M_center = s%M_center + s % dt/secyer * Mdotcore
call Coreradius(),
s%L_center =  3/5 * G *  s%M_center/s%R_center * Mdotcore
\end{verbatim}
where \verb!Mdotcore! indicates the fraction of the accreted heavy elements that are deposited onto the core. In the \verb!CoreRadius()! subroutine, we compute the core's radius using the tables of \cite{Chen2016} as described above.

\subsection{Gas accretion rate}
At each time-step we determine the gas (i.e., H-He)  accretion rate.
There are two different regimes for the gas accretion rate.
In the first, the outer planetary radius cannot be larger than either the  planet's Hill radius $R_H$ or the Bondi radius $R_B=GM_p/c_s^2$, where $c_s$ is the disk's sound speed at the planet's location, whichever is smaller (see \citealt{pollack1996} for details). 

In this first stage, when the proto-planet mass is low ($\sim \le 30$ M$_\oplus$) the gas accretion rate $\dot{M}_{xy}$ is obtained at each time-step through the requirement that the computed planet's radius $R_p$ equals to:
\begin{equation}
\label{radius}
	R_A=\frac{GM_p}{c_s^2+4GM_p/R_H}. 
\end{equation}
%where $c_s$ is the disk's sound speed at the planet's location and $R_H$ is the planet's Hill radius.
MESA is a lagrangian code where the independent coordinate is the mass. As a result, it is not possible to set the radius which is iteratively found at each time-step.
This is done by adding different amounts of gas until we reach convergence and $R_p$ matches  $R_A$ within a given precision.
This is done in the \verb!extras_check_model! routine. More details on the implementation of this routine are given in appendix \ref{Appendix -AttachedPhase}.

When the planet's mass is large enough so that the computed $\dot{M}_{xy}$ is greater than the amount of gas that the disk can supply ($\dot{M}_{xy,max}$), runaway gas accretion begins (detached phase). In this later stage the gas accretion rate is given by prescriptions based on hydro-dynamical simulations.
We implemented three different prescriptions: (1) the results of \cite{lissauer2009}, (2) the analytical result of \cite{Ginzburg2019}; and (3) Equation~16 of \cite{Tanigawa07}, where we substitute in $\Sigma_{acc}$ Equation~10 of \cite{Kanagawa17}. Table~\ref{table2} list the different gas accretion rates we consider.

\begin{table}[h!]
\begin{center}
 \begin{tabular}{c c c} 
 \hline
 Runaway gas accretion rate & Reference & $\alpha$\\ 
 \hline\hline
 $\dot{M}_{H-He}$1 & \cite{lissauer2009}, their Equation~3 & $4 \times 10^{-3}$\\ 
 $\dot{M}_{H-He}$2 & \cite{lissauer2009}, Equation~\ref{Eq.fitexpression}$^*$ & $4 \times 10^{-4}$ \\
 $\dot{M}_{H-He}$3 & \cite{Ginzburg2019}, their Equation~18 & $4 \times 10^{-3}$\\
 $\dot{M}_{H-He}$4 & \cite{Ginzburg2019}, their Equation~18  & $4 \times 10^{-4}$ \\
 $\dot{M}_{H-He}$5 & \cite{Kanagawa17}, their Equation~10 & $4 \times 10^{-3}$\\
 $\dot{M}_{H-He}$6 & \cite{Kanagawa17}, their Equation~10 & $4 \times 10^{-4}$\\
 \hline
\end{tabular}
\end{center}
\caption{The different (runaway) gas accretion rates used in this work. $^*$In the case of $\dot{M}_{H-He}$2, we fit the data derived by \cite{lissauer2009} (blue line in Figure~3 of their paper) with an high order polynomial. The coefficients of the fit can be found in Equation~\ref{Eq.fitexpression} of Appendix \ref{Appendix -DetatchedPhase}.}

\label{table2}
\end{table}

During this formation stage, the outer boundary conditions of the disk  \verb!fixed_T_and_P! are removed and  we adopt the \verb!simple_photosphere! MESA boundary condition. 
%the standard photospheric boundary condition of  MESA is adopted.
More details on the outer boundary conditions in this case can be found in the Appendix \ref{Appendix -DetatchedPhase}. 

\clearpage
\subsection{Planetesimal-envelope interaction}
\label{solidgasinteraction}
We include planetesimal's ablation and fragmentation in the calculation. The heavy-element mass that is deposited at each layer of the envelope is computed at each time-step following Equation~13 of VH19. 
The heavy elements in the model are represented by one substance which is taken to be water, and therefore we cannot properly model different types of heavy elements in the envelope in a self-consistent manner (e.g., 50 \% rock, 50 \% water). As a result, for the sake of simplicity,  in this work both planetesimals and pebbles are assumed to be composed of pure-water.  
Clearly it is desirable to include mixtures of heavy elements (e.g., water+rock) in the equation of state calculation, providing the possibility to include solids with various compositions and we hope to address this in future work. 

The heavy elements are added to the envelope by modifying the composition of the layers. We calculate the amount of heavy elements deposited at each layer and  increase the metallicity accordingly, \verb!Z(i) = Z(i) + deposited_mass / layer_mass!, where $Z(i)$ indicates the metallicity of the shell $i$.
The increased metallicity of the layer implies a decrease in the amount of H-He, which is then added at the top of the envelope using the \verb!mass_change! MESA variable. This could result in an artificial inflation of the radius and a decrease in the gas accretion rate. In appendix \ref{Appendix -AttachedPhase} we show that our results are insensitive to the metallicity of the outermost layer.

%Planetesimals are assumed to be composed entirely of water.  
The change in the envelope's composition has two main effects:
\begin{itemize}
	\item[(1.)] A change in the EOS.  
	The fraction of water determines the EOS at a given layer. This in turn changes the density, temperature, and pressure profile in the envelope. 
	\item[(2.)] A change in the opacity. 
	The  presence of heavy element affects the gas opacity which strongly affects the planetary growth \citep{Ikoma08,Mordasini14}.
	The opacity changes due to directs and indirect effects. A different metallicity leads to a direct change in the gas opacity.
	In addition, a different amount of heavy elements changes 
	the temperature and pressure in the envelope, which in turn affects the opacity.
\end{itemize} 

Fragmentation is the main mechanism for the mass deposition of large planetesimal ($r \geq 1$ km)  \citep{pollack96}, and is assumed to occur when the pressure gradient of the surrounding gas across the planetesimal exceeds the material strength (Equation~1 of VH19). We do not include Rayleigh-Taylor (RT) instabilities \citep{Mordasini2006,Korycansky00} to determine the fragmentation of the planetesimal. 
In appendix~\ref{section:materialstrength} we investigate the sensitivity of the results when changing the fragmentation strength and the latent heat of vaporization of the high-Z material. 

We include heavy-element settling following \cite{iaroslvaik07}. In this calculation, the heavy-element mass that settles to the underlying layers is determined by comparing the pressure of the heavy elements present in the layer to the water vapor pressure (the pressure above which water condenses). The vapor pressure of water is given by $P_{vap} = e^{-5640.34/T+28.867}$ (\textit{CRC Handbook of Chemistry and Physics}), where the temperature is in Kelvin and the pressure is in dyne cm$^{-2}$.  If the partial pressure exceeds the vapor pressure, the heavy elements are assumed to settle to the layer below. 
Energy is also added in each layer due to the conversion of the planetesimal's energy into  heat. The amount of energy $\Delta E_i$ released into a given atmospheric layer $i$ is given by: \citep{pollack86}
\begin{equation}
\label{added_energy}
	\Delta E_i = F_d ds_i + \frac{1}{2}\Delta m_i v_p^2,
\end{equation}
where $F_d$ is the drag force exerted on the planetesimal at layer $i$, $ds_i$ is the width of the layer $i$, $\Delta m_i$ is the mass of the planetesimal ablated in shell $i$, and $v_p$ is the planetesimal's local velocity. 
The first term on the right hand side of Equation~\ref{added_energy} represents the energy deposited in a layer due to gas drag that slows the planetesimal while the second term represents the deposition of the kinetic energy of the ablated material. The energy provided by the  planetesimals is included in MESA into the \verb!extra_energy! sub-routines.
If a fraction of the planetesimal's mass reaches the core, its energy is added onto the core's luminosity as:
\begin{equation}
\label{coreluminosity}
    L_{core} = \frac{GM_{core}}{R_{core}}\dot{M}_{core}, 
\end{equation}
where $\dot{M}_{core}$ is the total heavy-element mass that reaches the core.

\subsection{Heavy-element mixing}
\label{mixing}
After the heavy elements are deposited in the envelope we investigate whether they mix and redistribute by convection.  
The heavy-element distribution is affected by the mixing mechanism which is determined by heat transport mechanism which can be radiation, conduction, or convection.
To determine whether a region with composition gradients is stable against convection we use the standard Ledoux criterion $\nabla_{T} < \nabla_{ad} + (\varphi / \delta )\nabla_{\mu}$, where $\nabla_{T} = d \ln T / d \ln P$, $\nabla_{ad}$ and $\nabla_{\mu} = d \ln \mu / d \ln P$ are the temperature, adiabatic and the mean molecular weight gradient, respectively, while $\varphi = (\partial \ln \rho / \partial \ln \mu)_{P,T}$ and $\delta = (\partial \ln \rho / \partial \ln T)_{P,\mu}$ are material derivatives. 
In a homogeneous region, $\nabla_{\mu} = 0$ and the Ledoux criterion reduces to the Schwarzschild criterion.

If the mean molecular weight increases towards the planetary center the composition gradient acts against convection. Due to planetesimal fragmentation it can happen during the planetary growth that the mean molecular weight decreases towards planetary center, and the composition gradient leads to convective mixing. 
The mixing length theory (MLT) used in MESA requires the knowledge of a mixing length $l_m = \alpha_{mlt}H_P$ , where $H_P$ is the pressure scale-height and $\alpha_{mlt}$ is a dimensionless parameter. For stars the value of $\alpha_{mlt}$ is of the order of unity \citep{Sonoi19}. The value of $\alpha_{mlt}$ for planets is poorly constrained, but convincing arguments point towards values  significantly lower than for  stars \citep{Leconte12}. 
Following previous studies on Jupiter’s evolution with convective mixing \citep{Vazan15,Muller20} we adopt $\alpha_{mlt} = 0.1$. 
Since the efficiency of mixing does not play a big role in our study the exact value of $\alpha_{mlt}$ is of secondary importance. Nevertheless, we emphasize the importance of determining its value in the planetary regime.

\subsection{Putting the pieces together}
At time $t$ the planet has a total heavy-element mass $M_z$ and a total H-He mass $M_{xy}$. The outer radius is equal to $R_A$ given by Equation~\ref{radius}.

During a time-step $dt$ of the planet's growth we apply the following: 
\begin{enumerate}
    \item We compute the capture radius and the disk's solid surface density. We then  compute the solid accretion rate.
    \item Using the heavy-element accretion rate, we  compute the fraction that reach the core and the fraction that is dissolved in the envelope.  The heavy-element mass is deposited (non uniformly) in each layer following \ref{solidgasinteraction}.
    \item The planet is growing in mass and is heavy-element dominated. This has two effects: on one hand it enlarges the accretion radius $R_A$ given by Equation~\ref{radius}, on the other the planet's radius shrinks due to the increased gravitational mass and mean molecular weight, where $R_p \le R_A$.
    \item We compute the gas accretion rate in an iterative way. We solve the structure equations at each iteration assuming different H-He masses (i.e., H-He accretion rate) until we reach radius convergence, where $R_p \sim R_A$.

\end{enumerate}

\section{Results}

\label{sec:result}
\subsection{Formation without heavy-element enrichment}\label{sec:solidandgas}
For simplicity, in this section, we neglect the effect of heavy-element enrichment, and all the heavy elements are assumed to join the core. This assumption is relaxed in the following sections. 
We present Jupiter's formation history using different heavy-element accretion rates as well as various gas accretion rates for the runaway gas accretion phase.

First, in order to check the validity of our numerical methods, we present in Figure~\ref{Fig.10-Jupiter} a comparison between our work and Figure~12 of \cite{lissauer2009}.
We simulate the planetary formation assuming the solid accretion rate $\dot{M}_{Z}$1 of Table \ref{table1} and the runaway gas accretion rate $\dot{M}_{H-He}$2 of Table \ref{table2}. We call this run Z$_1$HHE$_2$, and in Figure~\ref{Fig.10-Jupiter} we compare it with simulation 3l$R_H$J of \cite{lissauer2009}.

\begin{table}[h!]
\begin{center}
 \begin{tabular}{c c c} 
 \hline
  Run & Heavy-element Accretion rate & Runaway gas accretion rate\\ 
 \hline\hline
 Z$_1$HHE$_2$ & $\dot{M}_{Z}$1  & $\dot{M}_{H-He}$2\\ 
 \hline
\end{tabular}
\end{center}
\caption{Solid and runaway gas accretion rate that we use in run Z1HHE2, compared with run 3l$R_H$J of \cite{lissauer2009}.}
\label{table3}
\end{table}

As can be seen from the figure there is a very good agreement.
The small differences between the two curves are probably caused by different assumed dust opacities.  

Figure~\ref{Fig.10-Jupiter} shows clearly the three phases of giant planet formation. Initially, the heavy-element accretion rate is very high, and the protoplanet reaches rapidly the planetesimal isolation mass ($< 1 Myr$). When the planet has accreted almost all of the solid mass present in its feeding zone (end of {\it phase 1}), the heavy-element accretion rate decreases and the proto-planet starts to accrete gas. This is known as {\it phase-2} or the {\it attached-phase} and its duration determines the formation time-scale. Finally, in {\it phase-3}, gas is accreted rapidly, the mass of the planet increases dramatically, and the planet opens a gap and detaches from the disk.

\begin{figure}[h]
	
	\centering
	
	\includegraphics[width=0.35\linewidth]{Jupiter.pdf}
	\caption{A comparison between our run Z$_1$HHE$_2$ (blue line) and run 3l$R_H$J of \cite{lissauer2009}  (orange line).}
	\label{Fig.10-Jupiter}
\end{figure}

\begin{figure}[h!]
	\centering
	\includegraphics[width=0.625\linewidth]{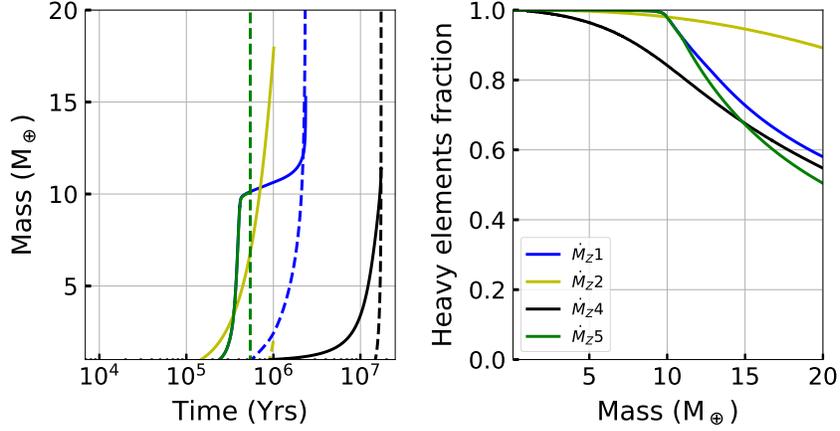}
	\caption{\textbf{Left:} Jupiter's growth assuming different heavy-element (solid) accretion rates. The dashed lines indicates the H-He mass  while the solid lines represent the amount heavy elements that are assumed to join the core. The different colors represent different heavy-element accretion rate. The blue, yellow, black and green  lines correspond to heavy-element accretion rate $\dot{M}_{Z}$1, $\dot{M}_{Z}$3, $\dot{M}_{Z}$4, $\dot{M}_{Z}$5. 
		\textbf{Right:} The planetary bulk metallicity (Equation~\ref{bulk}) as a function of the total planetary mass. The colours scheme is the same as for the left panel.}
	\label{Fig.6-Solidaccretionrate}
\end{figure}

We next compare Jupiter's growth assuming different accretion rates. The planet is assumed to be formed \textit{in situ}, i.e., at 5.2 AU. The results are summarized in Figure~\ref{Fig.6-Solidaccretionrate}.
The left panel shows the growth of the planetary mass as a function of time assuming different solid (heavy-element) accretion rates. 
The blue curve represents the solid accretion rate $\dot{M}_{Z}$1 \citep{pollack1996}.  
Initially, the planetesimals accretion rate is very high, and isolation mass ($\sim$ 10 M$_\oplus$) is reached rapidly. Then, the solid accretion rate decreases and gas starts to be accreted efficiently. Finally, runaway gas accretion occurs and the planet detaches from the disk.
The solid accretion rate $\dot{M}_{Z}$4 \citep{Fortier07} is indicated by the black line. This accretion rate results in the longest formation timescale , which is substantially larger than typical lifetimes of protoplanetary disks \citep{Mamajek09,Pfalzner2014,Ribas15}. 
Note that in the oligarchic growth scenario, the planet does not grow via the standard three phases.  The oligarchic regime rate predicts a low accretion rate at the beginning of the simulation which is slowly growing with time, and as a result, the feeding zone is never depleted (no phase-2). Therefore, in this scenario there is no isolation mass.
The green line corresponds to the solid accretion rate $\dot{M}_{Z}$5 \citep{Shiraishi08}. 
In this case the gas accretion rate is already very high after that the planet reaches $\sim$ 10 M$_\oplus$. Before the planet reaches $\sim$ 10 M$_\oplus$ we use the solid accretion rate of $\dot{M}_{Z}$1. In this scenario, a giant planet is formed very rapidly, in a time-scale shorter than a million years.

Finally, the yellow line corresponds to the pebble accretion rate $\dot{M}_Z$2 \citep{Lambrechts14}. At Jupiter's location the pebble isolation mass is $\sim$ 20 M$_\oplus$. In this scenario, the formation of a giant planet is much shorter than in the case of planetesimal accretion, as already indicated by previous studies \citep{Lambrechts14,VenturiniHelled17,Lambrechts17}. 

The right panel of Figure~\ref{Fig.6-Solidaccretionrate} shows the planetary bulk metallicity as a function of the total planetary mass, which is defined as: 
\begin{equation}
\label{bulk}
Z_{bulk} = \frac{M_Z}{M_Z + M_{H-He}},
\end{equation}
where M$_Z$ is the total heavy-element (solid) and M$_{H-He}$ is the H-He mass.
In the standard oligarchic planetary growth $\dot{M}_{Z}$4, the planet's metallicity decreases almost linearly with planetary mass. 
The prescription for $\dot{M}_{Z}$5 \citep{Shiraishi08} results in very efficient H-He accretion leading to a low bulk metallicity during all the stages of the planetary formation. 
For the case of pebble accretion with $\dot{M}_{Z}$2 it is found that gas accretion is inefficient and the bulk planetary metallicity remains very high. 
Finally, assuming the standard solid accretion rate of \cite{pollack1996} ($\dot{M}_{Z}$1) leads to a bulk metallicity close to one (i.e., mostly heavy elements) until the feeding zone is depleted and gas is accreted.  At this point, the planetary metallicity decreases rapidly.

In each case the gas accretion is computed via the requirement that the planet's radius $R_p$ matches $R_A$ introduced in Equation~\ref{radius}. During this stage, the proto-planet accretes H-He gas due to cooling. At each time-step, the planet radiates energy, cools down, and therefore can accrete more gas. It is clear from Figure~\ref{Fig.6-Solidaccretionrate} that H-He accretion is driven by the heavy-element accretion rate. Different solid accretion rates result in different H-He accretion rates.

At some point the disk is unable to provide gas at a sufficient rate, and the computed $\dot{M}_{gas}$ exceeds the maximum amount of gas that the disk can supply. The gas accretion is then controlled by hydrodynamics rather than thermodynamics. During this phase the planet detaches from the disk. The outer boundary conditions of the disk  \verb!fixed_T_and_P! are removed and the \verb!simple_photosphere! option of  MESA is adopted. Therefore, the surface temperature, pressure, and density are changed and evolve with time. More details can be found in Appendix \ref{Appendix -DetatchedPhase}.

We investigate the differences in the inferred planetary radius, luminosity, and mass when we use  different H-He accretion rate (as listed in Table~\ref{table2}).

Figure~\ref{Fig.8-RunawayFormuale} shows Jupiter's growth when we use the different runaway gas accretion formulae assuming two different disk $\alpha$ viscosity parameter of $3 \times 10^{-3}$ and $3 \times 10^{-4}$, indicated by solid and dashed lines, respectively. 
For the purpose of comparison, in all the cases the initial mass is set to $0.5$ M$_J$. All the accretion rates are multiplied by a cap function equals to 
\begin{equation}
	f_{cap} = 1-\frac{M_p}{M_J},
\end{equation}
which has the role to terminate the planet's growth smoothly at Jupiter's mass.

The three different formulae predict that the gas accretion rate decreases with increasing viscosity.
The formulation of \cite{Kanagawa17} ($\dot{M}_{H-He}$5 and $\dot{M}_{H-He}$6) predicts the fastest planetary growth. 
The planet's growth based on the runaway gas accretion rate $\dot{M}_{H-He}$1 \citep{lissauer2009} is in excellent agreement with the  $\dot{M}_{H-He}$5 case  \cite{Kanagawa17}, but there is a clear difference when we compare $\dot{M}_{H-He}$2 to $\dot{M}_{H-He}$6. The analytical result of \cite{Ginzburg2019} (runs  $\dot{M}_{H-He}$3 and  $\dot{M}_{H-He}$4) leads to a slow growth of the planet, and when coupled with a low value of viscosity, it results in a rather long formation timescale ( $> 10$ million years). 
The middle panel of the figure shows the planet's radius as a function of planetary mass for the different formulae. It is clear from the figure that the exact value of the planetary radius changes significantly, depending on the assumed accretion rate.
The differences can be up to several Jupiter radii. 
The cases associated with faster planetary growth result in a larger radius. This happens because for these cases the growing planet has less time to cool down. 
Finally, the right panel presents the planetary luminosity as a function of mass. As the radius, also here there is a substantial difference up to several orders of magnitude when the different prescriptions are used. 
In this phase, the planet's luminosity is driven by the gas  accretion the planetary contraction, which cause a change in the gravitational energy. Therefore, different gas accretion formulae result in different luminosities.
Determining the planetary luminosity of young giant planet is of significant importance for determining the masses of planets detected by direct imaging, especially of young planetary candidates that are still accreting gas from their proto-planetary disk \citep{Sallum15,Guidi18,Wagner18}. 
The luminosity of such planets is a key observable; it provides hints about the thermodynamic state of the planet \citep{Berardo17} and it can be used to constrain planet formation theories (e.g. \cite{Marleau13}).
%they constrain mass accretion rates and the thermodynamical state and planet formation theories by.
This highlights the importance of accurately modeling the terminal phase of nebular accretion, and we suggest that this topic should be investigated in more detail in future studies.

\begin{figure}[h]
	
	\centering
	
	\includegraphics[width=0.65\linewidth]{Runawaygasaccretion.pdf}
	\caption{\textbf{Left:} The total planet's mass as a function of time for three different runaway gas accretion formulae. The orange, blue and black lines correspond to the prescriptions of \cite{lissauer2009}  ($\dot{M}_{H-He}$1 and $\dot{M}_{H-He}$2) , \cite{Ginzburg2019} ($\dot{M}_{H-He}$3 and $\dot{M}_{H-He}$4)  and \cite{Kanagawa17}  ($\dot{M}_{H-He}$5 and $\dot{M}_{H-He}$6) , respectively. Solid lines correspond to  disk's alpha viscosity parameter of $3 \times 10^{-3}$, while the dashed ones represent simulations in which the disk's alpha viscosity is assumed to be $3 \times 10^{-4}$. \textbf{Middle:} Planetary radius, in Jupiter radii, as a function of planetary mass for the different runaway formulae. \textbf{Right:} Planetary luminosity, in solar units.}
	\label{Fig.8-RunawayFormuale}
\end{figure}

\newpage
\subsection{Planetesimal accretion}
\label{sec:planetesimals}

In this section, we include heavy-element enrichment in the planetary envelope when modeling Jupiter's formation. 
We use the numerical tools described in section \ref{sec:methods} to include the deposition of mass and energy of 100 km-sized planetesimals. 

As already shown in VH19 the main mass deposition mechanism of a large planetesimal is fragmentation. 
The planetesimal's velocity increases as it travels towards the planetary center until the pressure acting upon its surface is greater than the material strength and leading to fragmentation. 
The details on the fragmentation process and the distribution of the material are important \citep{Valletta19,register17,Revelle05,Hills93}. 
%Fragmentation can be instantaneous, where all the planetesimal's mass is deposited into the envelope layer; 
%{\bf fix -- this still needs work... it is too much bla bla...}
%the planetesimal can be divided into two planetesimals of equal mass that will continue to travel towards the core, or it can be transformed into a cloud of heavy-element material.
Different fragmentation models result in different distributions for the heavy elements which in return also affect the inferred core mass and further growth. 

\begin{figure}[h!]
	\centering
	\includegraphics[width=0.4\linewidth]{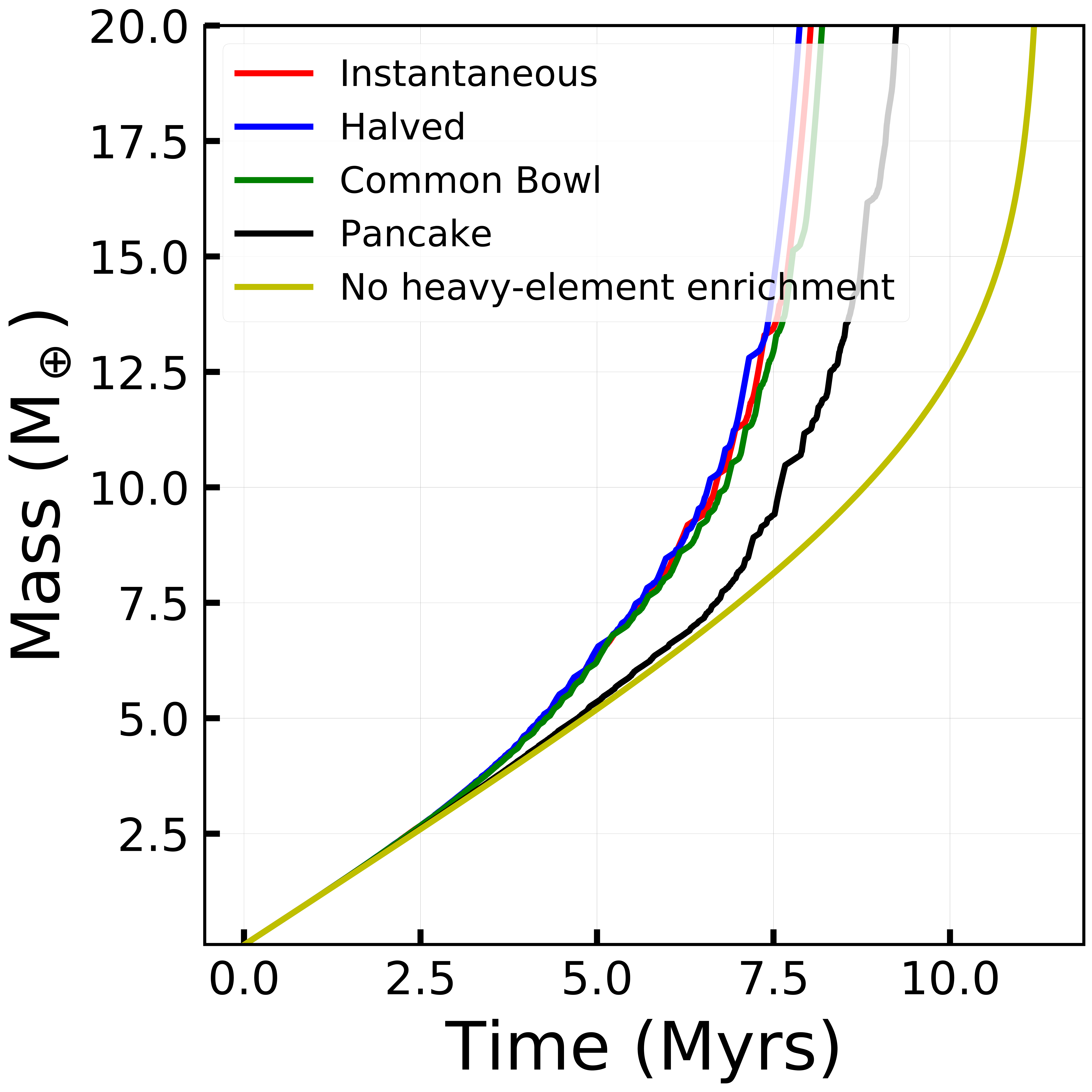}
	\caption{Total planetary mass as a function of the time for the different assumed fragmentation models.}
	\label{fragmentation_model}
\end{figure}

%As we discuss in VH19 different fragmentation models result in different core masses and heavy-element  distributions.
Figure~\ref{fragmentation_model} shows the total planet's mass vs.~time when using  different fragmentation models (see VH19 for details).  
It can be seen that instantaneous, common bowl and halved model lead to a similar planetary growth history, while the pancake model results in a slower formation time. 
The yellow line corresponds to a case where all the heavy elements are assumed to join core. 
%The envelope's composition, in this case, does not change with times and is set to  proto-solar. 
In this case the envelope's composition does not change and is set to $0.02$.
The different derived formation timescales when using the various fragmentation models is affected by the mass of the inner compact core.  
As shown analytically by \cite{Brouwers2017} the planetary growth timescale when heavy-element enrichment is considered depends on the heavy-element fraction that is deposited into the envelope relative to the amount that is added to the core. 
The smaller the inner core is, the faster is the growth of the planet.
This relation can be seen clearly from Figure~7 of VH19. The instantaneous, common bowl, and halved fragmentation models lead to very similar (compact) core masses, while the pancake model results in a more massive core. 
%allows for a much larger inner core, because the cloud of dust formed after fragmentation is difficult to be fully ablated.

\begin{figure}[h]
	\centering
	\includegraphics[width=0.65\linewidth]{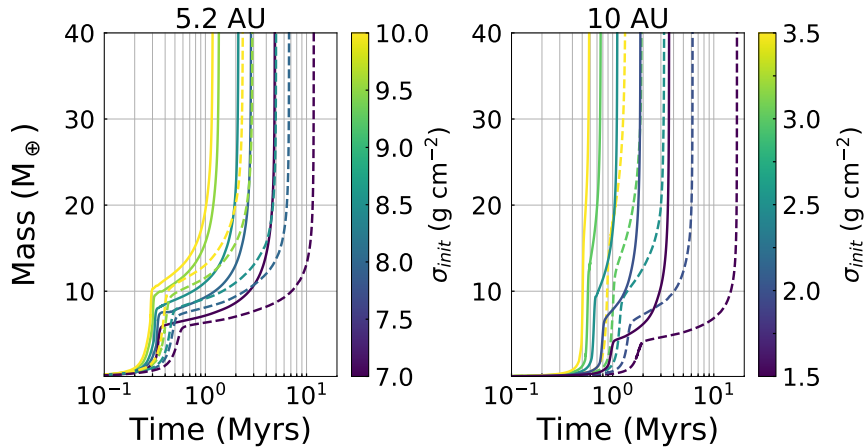}
	\caption{\textbf{Pollack+96 case:} Total planetary mass vs.~time for different assumed values of the initial solid surface density $\sigma_{init}$ represented by different colors. The solid lines represent cases where the heavy-element enrichment is included, while the dashed ones correspond to  cases where all the heavy elements are assumed to join the core. The right and left panels correspond to formation locations of 5.2 AU and 10 AU, respectively. } 
		\label{Fig-Pollack}
\end{figure}

Figure~\ref{Fig-Pollack} shows the effect of including heavy-element enrichment assuming the standard core accretion rate of \cite{pollack1996}.

We use the solid accretion rate  $\dot{M}_{Z}$1, but but investigate how the planetary growth depends on the assumed solid surface density.
The assumed initial solid surface density $\sigma_{init}$ in the planet's feeding zone is a key property that influences the formation timescale and growth history.
The dependence of the planet's growth on the initial solid surface density is consistent with the results of \cite{pollack1996}, where it was shown that the formation time-scale goes as $\sim \sigma_{init}^{-4.5}$.
We assume different solid surface densities ranging between 7 and 10 g cm$^{-2}$ as indicated by the different colors. For {\it in situ} formation of Jupiter the value of 10 g cm$^{-2}$ corresponds to three times the density of the minimum mass solar nebula model. 
The solid curves in Figure~\ref{Fig-Pollack} correspond to the case where heavy-element enrichment is included, while the dashed ones represents cases where all the heavies are assumed to reach the core. 
It is clear that the formation timescale is significantly reduced when including planetesimal ablation and the enrichment of the envelope.  
The increased mean molecular weight leads to more efficient gas accretion and, as a result, the formation timescale is shortened \citep{Stevenson82, Venturini16}.
When heavy-element enrichment is neglected, the formation timescale for the case with $\sigma_{init} = 7$ g cm$^{-2}$ exceeds the expected lifetimes of protoplanetary disks. 
However, with heavy-element enrichment, this relatively low solid surface density results in a formation timescale that is compatible with disks lifetime \citep{Ribas15,Pfalzner2014}.

For comparison, we also consider a case where the planet is formed at 10 AU (corresponding to Saturn's location). The results are shown in the right panel of Figure~\ref{Fig-Pollack}. 
Since the solid surface density decreases with increasing radial distances, as well as the disk's temperature and pressure, the growth rate and outer boundary conditions for the simulations are modified.
As expected, for Saturn's location the formation timescale is significantly longer. 

\begin{figure}[h!]
	\centering
	\includegraphics[width=0.65\linewidth]{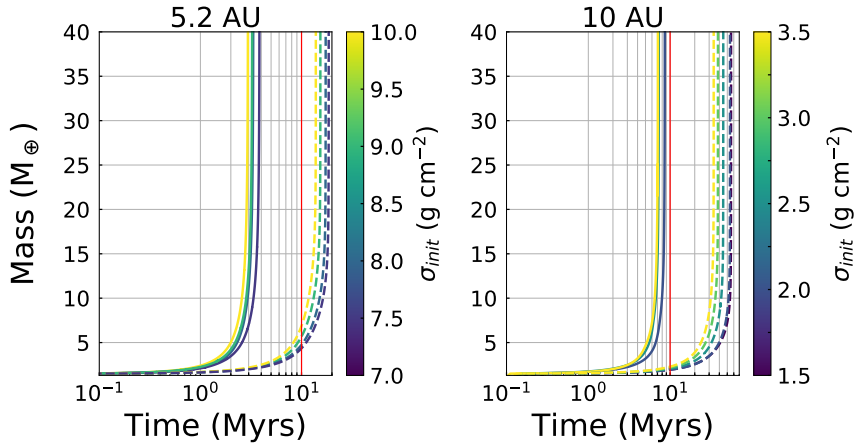}

\caption{{\bf Oligarchic growth case:} Total planetary mass vs.~time for different assumed values of the initial solid surface density $\sigma_{init}$ represented by different colors. The assumed solid accretion rate is taken from \cite{Fortier07}. The solid lines represent cases where the heavy-element enrichment is included, while the dashed ones correspond to  cases where all the heavy elements are assumed to join the core. The right and left panels correspond to formation locations of 5.2 AU and 10 AU, respectively. 
The red line represents the typical maximum disk lifetime, which is assumed to be 10 Myrs.
}
 \label{Fig-Fortier}
\end{figure}

\newpage
\subsection{Oligarchic regime}
\label{Sec-oligarchic}
In this section, we investigate the effect of heavy-element enrichment when assuming that the planet growth follows the oligarchic regime ($\dot{M}_Z$4).
Figure~\ref{Fig-Fortier} is similar to Figure~\ref{Fig-Pollack} but using the heavy-element accretion rate derived by \cite{Fortier07}. We indicate a maximum disk's lifetime by a red line, which is placed at 10 Mys. 
If all the solids join the core and heavy-element enrichment is neglected, the oligarchic regime is incompatible with formation timescales of a few Mys  \citep{Fortier07}, as can be seen from the figure.
%As already discussed in \cite{Fortier07}, the oligarchic regime is incompatible with formation timescales of a few Myr, as can be seen from the figure. 
For both assumed locations and for all the assumed solid surface densities, the dashed lines are beyond the red line, suggesting that the disk dissipates before the planet reaches runaway gas accretion. 
Furthermore, in this case the planetary growth is less sensitive to the initial solid surface density. 
As already seen in Figure~\ref{Fig-Pollack} including heavy-element enrichment significantly shortens the formation timescale. When including heavy-element enrichment, even the low solid accretion rate of the oligarchic regime leads to a  formation time-scale that is compatible with the disk's lifetime. In the left panel which corresponds to a formation location of 5.2 AU, all the solid lines reach 40 M$_\oplus$ within 3 - 4 Myr.
We conclude that the oligarchic accretion rate, coupled with heavy-element enrichment, can realistically be applied for giant planet formation at Jupiter's location. 
For a formation location of 10 AU, including heavy-element enrichment leads to formation timescales between seven and ten Myr (see right panel).
Overall, as expected, the planetary growth is faster at 5.2 AU than at 10 AU, where in the latter formation location the growth timescale is comparable to the disk's lifetime. 
This difference in the formation timescale could explain the different masses of Jupiter and Saturn. We suggest that this topic should be investigated in more detail and we hope to address this in future research.
%7-10 millions of years. At this point, probably, the disk has already been dissipated. 
\clearpage
\subsection{Pebble Accretion}
\label{sec:pebbles}
\begin{figure}[h!]
	\centering
	
	\includegraphics[width=0.65\linewidth]{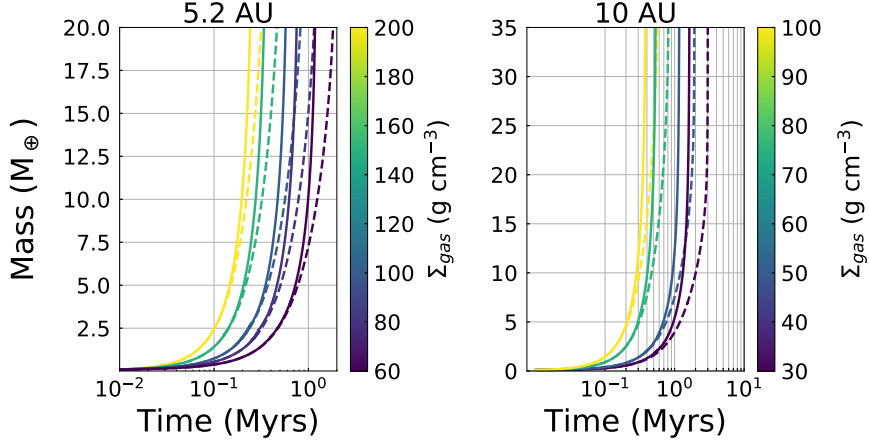}
	\caption{\textbf{Pebble accretion case:} 
	Total planetary mass vs.~time for growth via pebble accretion assuming  different values of the gas disk's density $\Sigma$ in the feeding zone represented by different colors. The solid lines represent cases where the heavy-element enrichment is included, while the dashed ones correspond to  cases where all the heavy elements are assumed to join the core. 
	The right and left panels correspond to formation locations of 5.2 AU and 10 AU, respectively.}
	\label{Fig.12-Pebblesformation}
\end{figure}

In this section, we investigate the effect of heavy-element enrichment on Jupiter's formation assuming that the planet is growing via pebble accretion. 
We use the solid accretion rate $\dot{M}_{Z}$2 but we perform a parameter study on the gas surface density.
Figure~\ref{Fig.12-Pebblesformation} shows the planetary growth as a function of time at two different locations: $5.2$ and $10$ AU, for various gas surface densities, with and without heavy-element enrichment.

As expected, the planetary formation with pebble accretion is faster and more efficient compared to the planetesimals' case. There are no runs where the formation timescale exceeds the expected disk's lifetime. When heavy-element enrichment is included, for a large set of parameters, the predicted time-scale is less than 1 Myr, which is rather too fast and is in conflict with the occurence rates of giant exoplanets  \citep{Venturini16,VenturiniHelled17}.
However, as shown by various studies \citep{Ormel15,Alibert15} a significant fraction of the accreted pebbles can be recycled back onto the disk. This can lead to a slower growth of the planet, reducing the efficiency of giant planet formation in the pebble accretion scenario.  
It is therefore desirable to investigate heavy-element enrichment in the pebble accretion model where pebble recycling is included. 
%should be investigated in the future. }

Pebble accretion is characterized by the so called pebble isolation mass. This is the mass at which the perturbation of the planet on the disk causes the pebble accretion rate to stop. It is typically assumed \citep{Lambrechts17,Lambrechts14,Bitsch15} that the proto-planet begins to accretes substantial amounts of gas only  after pebble isolation mass is reached, while before that it is {\it assumed that the gaseous (H-He) mass is 10 \% of total planetary mass}.

\begin{figure}[h!]
	\centering
	\includegraphics[width=0.55\linewidth]{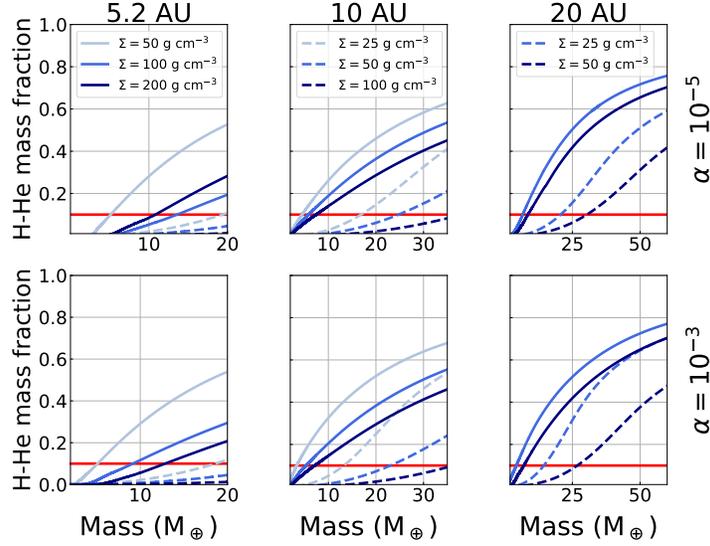}
	\caption{H-He mass fraction for different planetary locations and disk's viscosity. Each column  corresponds to a different formation  location, while each line indicates  a different disk viscosity. Solid lines indicate runs where heavy-element enrichment is included, while the dashed lines correspond to cases where solids are assumed to join the core. The red line indicates the assumption of \cite{Lambrechts17} and \cite{Bitsch15} where the H-He mass fraction is 10\% of the total mass until pebble isolation mass is reached. All the plots end at the pebble isolation mass.}
		\label{Fig.11-Pebblesfraction}
\end{figure}

Figure~\ref{Fig.11-Pebblesfraction} shows the inferred H-He fraction for the different cases, including (solid line) and neglecting (dashed line) heavy-element enrichment. 
The different columns represent different formation locations, while different line styles indicate different disk's $\alpha$ viscosity. The red line indicates a H-He mass fraction of 10\% as assumed in the pebble  accretion simulations. %it is typically assumed in different works.
We find that the fraction of H-He is increasing as pebbles are accreted. Therefore, assuming that the H-He mass fraction is constant until pebble isolation mass is reached is an over-simplification. Nevertheless, in some cases the fraction of H-He is indeed around 10\% when pebble isolation mass ($\sim$ 20 M$_\oplus$) is reached. 
We suggest that this approximation is appropriate for formation location of 5.2 AU and for small values for the the disk's viscosity.
However, at 10 AU and 20 AU the pebble isolation mass is significantly higher, being 35 and 60 M$_\oplus$, respectively. In these cases we find that the growing  protoplanet accretes large amounts of H-He of the order of several tens of percents.  Again, when  heavy-element enrichment is included gas can be accreted more efficiently.

%\clearpage
%\newpage
\section{The connection between the accretion rates and the planetary internal structure}
\label{sec:planetarymemory}

The heavy-element distribution within the planetary envelope has an important impact on the thermal evolution and on the final planet's structure \citep{Vazan15,Vazan17}. 
%Therefore, it is important to infer the primordial heavy-element profile from simulations.
In this section, we provide an approximation for the heavy-element distribution in the planet's envelope at the end of phase-2, i.e.,  before runaway gas accretion begins. This approximation can be used to infer the heavy-element profile of the deep interiors of gas giants or the structure of  intermediate-mass planets ($\sim10$ - $50$ M$_\oplus$) that never reached runaway gas accretion, such as Uranus and Neptune. We show that the heavy-element profile of such planets mainly depends on the ratio between the heavy-element and gas accretion rates. 

At the beginning of the planetary formation, the envelope's mass is negligible and the heavy elements deposit their mass in the deep interior. Once the envelope's mass becomes a significant fraction of the total mass, ablation becomes important and the heavy elements are deposited far from the planetary center.
For simplicity, we assume that heavy elements are deposited in the outermost layer. This is a reasonable approximation in particular for the case of water (ice) planetesimals which are easily fragmented or vaporised in the upper  envelope, and therefore typically deposit their mass in the outer regions.  
Rocky planetesimals can penetrate deeper into the envelope and the approximation becomes less accurate. 
For the case of icy planetesimals, the planetary structure can be viewed as an onion of different shells with different metallicities, where each of them is identified by its mass coordinate $m$ and corresponding metallicity associated with the accretion rates \cite{Stevenson82, HelledStevenson17}.

%If the heavy elements are assumed to be deposited in the outermost layer, then the planetary structure can be viewed as an onion of different shells with different metallicities, where each of them is identified by its mass coordinate $m$ and corresponding metallicitiy associated with the accretion rates \cite{Sttevenson, HelledStevenson17}. 
If the heavy elements remain where they are deposited (i.e., no convective mixing) then the metallicity $Z$ of a particular shell $i$ reflects the solid (heavy-element) and gas accretion rate at the time when that shell was the outermost layer. 
In this case the heavy-element profile as a function of the mass coordinate $z(m)$ can be approximated by \citep{HelledStevenson17}:
\begin{equation}
\label{heavyapproximation}
Z(m) \sim \frac{\dot{M}_{Z,env}(M)}{\dot{M}_{xy}(M)+\dot{M}_{Z,env}(M)},
\end{equation}{}
where $M_{Z,env}$ is the heavy-element mass dissolved in the envelope, which can be computed following the semi-analytical formula presented in VH19, and ${M}_{xy}$ is the H-He mass.  
On the left-hand side of Equation~\ref{heavyapproximation} $m$ represents the mass coordinate in the profile while on the right-hand side $M$ is the total planetary mass.

\begin{figure}[h!]
	\centering
	\includegraphics[width=0.42\linewidth]{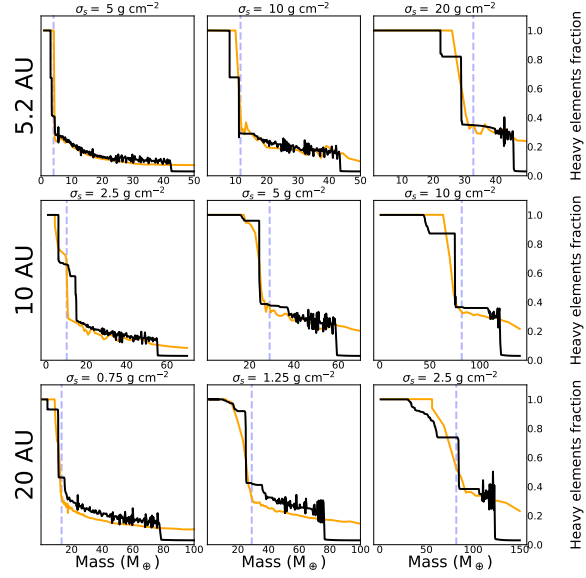}
	\caption{\textbf{Growth via planetesimal accretion:} Comparison between the accretion ratio (Equation~\ref{heavyapproximation}) and the heavy-element profile for 9 cases, each with a different location and initial solid surface density. Each line represents a different location in the disk of the proto-planet. The blue dashed line represents the planetesimal isolation mass.}
	\label{Planetesimal-Approximationwithreal}
\end{figure}

\begin{figure}[h]
	
	\centering
	
	\includegraphics[width=0.42\linewidth]{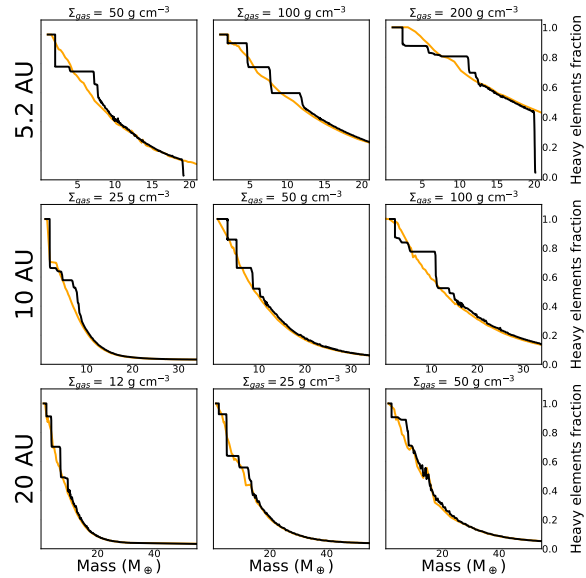}
	\caption{\textbf{Growth via pebble accretion:} Comparison between the accretion ratio (Equation~\ref{heavyapproximation}) and the heavy-element profile for 9 cases, each with a different location and initial solid surface density. Each line represents a different location in the disk of the proto-planet.}
	\label{Pebbles-Approximationwithreal}
\end{figure}

We test this approximation for the cases of both planetesimal and pebble accretion assuming different heavy-element accretion rates. The results are presented in Figures  \ref{Planetesimal-Approximationwithreal} and \ref{Pebbles-Approximationwithreal}, respectively. 
The orange line corresponds to Equation~\ref{heavyapproximation} while the black one indicates the inferred heavy-element profile in the planet's envelope. 
%Where the heavy-element mass fraction $Z$ is defined by $ Z = {M_z}/{dm}$, 
The calculation includes convection (using the Ledoux criterion) and the sinking of heavy-element as described in section~\ref{sec:methods}. 
Therefore, Equation~\ref{heavyapproximation} does not always perfectly reproduce the $Z$ profile, since convective mixing redistributes the heavy elements. However, the orange line follows qualitatively the behaviour of the black one.

Figure~\ref{Planetesimal-Approximationwithreal} corresponds to the case of planetesimal accretion using $\dot{M}_Z$1. The solid surface density and the planet's location are specified in the figure.
%approximation \ref{heavyapproximation} for planetesimals. 
As discussed above, large planetesimals deposit mass in the envelope mainly with fragmentation.
Fragmentation creates jumps in the planet's heavy-element profile due to the deposition of all the planetesimal's mass in a small location in the envelope.
As a result, there is a difference between the approximation and the actual profile of the planet, but still the orange line qualitatively follows the the black one for all the assumed locations solid surface densities.
The blue dashed line represents the planetesimal isolation mass. For M $\lesssim$ M$_{iso}$, the heavy-element fraction is high and $Z \sim 1$. Once the planetesimal isolation mass is reached, the heavy-element fraction decreases rapidly.

The case of pebble accretion ($\dot{M}_Z$2) is presented in Figure~\ref{Pebbles-Approximationwithreal}. 
The location of the planet and the gas surface density are specified in the figure.
%the proto-planet is assumed to accrete pebbles.  
It can be seen that in all plots, there is a convective inner region in the envelope, where the red and black line do not overlap.  In the outer, radiative part, the agreement between the planetary profile and Equation~\ref{heavyapproximation} is excellent.

The approximation given by Equation~\ref{heavyapproximation} can be used to investigate the \textit{primordial} heavy-element profile in the deep interior of giant planets or the overall profile of  intermediate-mass gas-rich  planets, ($\sim$10 - 50 M$_\oplus$). 
It should be noted, however, that this calculation corresponds to the planetary primordial internal structure which could change during the planetary long-term evolution due to due to convective mixing. Nevertheless, our calculation implies that the deep interiors of Uranus and Neptune shortly after their formation are expected to have composition gradients. 
%The heavy-element profile present today in such planets, Uranus and Neptune for example, can be different from what inferred by Eq.~\ref{heavyapproximation} due to convective mixing in  the long term evolution.}

\clearpage
\section{Summary and Conclusions}
\label{conclusion}
We developed a new numerical framework for giant planet formation using MESA. 
We include the dissolution of the accreted heavy elements self-consistently, where the high-z material is included in the EOS and opacity calculations. 
Our formation model can be used to model the deposition of the high-Z material in the envelope of growing planets and to follow the planetary long-term evolution.

We investigated Jupiter's growth history assuming  different solid (heavy-element) and runaway gas accretion rates, different sizes of the accreted solids,  and fragmentation model. 
We find that the planetary growth strongly depends  on the assumed solid and runaway gas accretion rates. Very different formation time-scales are inferred when assuming different solid accretion rates. For example, when heavy-element enrichment is neglected, the oligarchic regime leads to formation time-scales longer than 10 Myrs (the maximum expected disk's lifetime). 
This is not the case when assuming the solid accretion rate from \cite{pollack1996}.

In addition, the planetary luminosity shortly after formation depends on the assumed runaway gas accretion rate. Different runaway gas accretion rates can result in a difference of three order of magnitude in the planetary luminosity (!).
This should be taken into account when determining the masses of giant  planets detected by direct imaging, since a given planetary mass can have different luminosity depending on its growth history (runaway accretion). This is particularly relevant for young planetary candidates that are still accreting gas from their proto-planetary disk \citep{Sallum15,Guidi18,Wagner18}.
As a result, the terminal phase of gas accretion should be investigated in more detail.

In all the simulations we performed including heavy-element enrichment lead to shorter formation time-scales. This has been already shown by different authors (e.g., \citealt{Stevenson82, Venturini16}) and is a consequence of the higher mean molecular weight of the envelope, which results in a more efficient accretion of gas.
%We confirm that including heavy-element enrichment in the calculation of the planetary growth {\it always} leads to shorter formation time-scales \citep{VenturiniHelled17,pollack86,Brouwers20} due to a more efficient gas accretion.  
For the case of pebble accretion, we find that the fraction of H-He in the planet can significantly exceed 10\% even before pebble isolation mass is reached. 
We also show that when we assume the oligarchic accretion rate from \cite{Fortier07} including heavy-element enrichment, the formation time-scale is reduced to a few Myr, a timescale which is consistent with the estimated lifetimes of protoplanetary disks.

Finally, we also show that the heavy-element profile within the planet's deep interior (before runaway gas accretion) reflects its accretion history. i.e., the ratio between the heavy-element and gas accretion rates. This, however, assumes that no significant mixing has occurred during the planetary evolution.
Indeed, recent structure models of Jupiter suggest that the planet has a fuzzy core \citep{wahl2017,Debras19}. Such a structure could be a result of primordial composition gradients that do not mix during the planetary evolution \citep{Vazan17}. However, it was recently shown by \cite{Muller20} that composition gradients associated with the formation process are unlikely to retain their primordial shape. It was shown that Jupiter's envelope after runaway gas accretion is hot enough for the majority of the envelope to become convective, despite the stabilizing composition. Therefore, the origin of Jupiter's extended diluted core is still unknown and should be investigated further. 

%The results presented here are not only relevant for understanding the cores of giant planets, but also the formation history and internal structures of Uranus and Neptune and exoplanets in this mass/size range. 

Our study represents an additional step forward in giant planet studies, but clearly much more work is needed. 
In this work, the high-Z material was represented by water. Although the results are insensitive to the assumed material strength and density of the accreted planetsimals as shown in Appendix D., assuming a different material (e.g., rock) would affect the EOS and opacity calculations, and therefore the planetary growth. 
As a consequence, including the presence of other elements in the EOS and opacity calculations is desirable as it can impact the growth, internal structure and long-term evolution of the planet \citep{Vazan13}. 
Furthermore, a more realistic treatment for the dust opacity with a proper modeling the grain microphysics is desirable. We hope to address these topics in future research.
Future investigations should also include a hybrid pebble-planetesimal formation path \citep{Venturini19,Yann18}. 
Finally, using this new numerical framework,   future studies could focus on the 
formation history, evolution, and internal structures of Saturn, Uranus and Neptune, as well as gaseous-rich exoplanets. 

%could focus on the formation and subsequent evolution of Saturn, Uranus and Neptune, as well as gaseous-rich exoplanets. 

\subsection*{Acknowledgments}
We thank Simon M{\"u}ller and Andrew Cumming for valuable discussions and technical support.
We also acknowledge Julia Venturini for sharing her results with us.   
We thank the anonymous reviewers for their careful reading of our manuscript and their insightful comments.
RH acknowledges support from the  Swiss National Science Foundation
(SNSF) via grant 200020\_188460.

\clearpage
\appendix
\section{Comparison with standard assumptions}
The stellar structure equations of mass conservation, hydrostatic balance, thermal gradients, and energy conservation are described by:
\begin{equation}
\frac{dm}{dr} = 4\pi r^2 \rho
\end{equation}
\begin{equation}
\frac{dP}{dr} = -G\frac{m}{r^2}\rho 
\end{equation}
\begin{equation}
\frac{dT}{dr} =  \nabla \frac{T}{P}\frac{dP}{dr}
\end{equation}
\begin{equation}
\label{dldr}
\frac{dL}{dr} = 4\pi r^2 \rho\bigg(-T\frac{dS}{dt}\bigg), 
\end{equation}
where $r$ is the radial coordinate, $L$ is the luminosity, $T$, $P$, $\rho$ and $S$  are the temperature, pressure, density and entropy, respectively.  $\nabla$ is the temperature gradient which depends on the heat transport mechanism within the envelope. 
In many planetary formation/evolution codes \citep{Piso14,Piso15,Venturini16,VenturiniHelled17} Equation~\ref{dldr} which describes the local conservation of energy is replaced by a global energy balance. 
The set of time-dependent partial differential equations (PDEs) describing the evolution of the planet is transformed into a set of time independent ordinary differential equations (ODEs), which is more easy to be solved numerically. 
However this transformation has two main limitations. First, these codes cannot be used to properly model the planetary long-term evolution of the planet because the time is not present anymore in the equations. Second,  Equation~\ref{dldr} is no longer satisfied, and the planetary luminosity is assumed to be constant within the envelope. 

Below we explore the limitations of this approach. 

\begin{figure}[h!]

	\centering
	\includegraphics[width=0.65\linewidth]{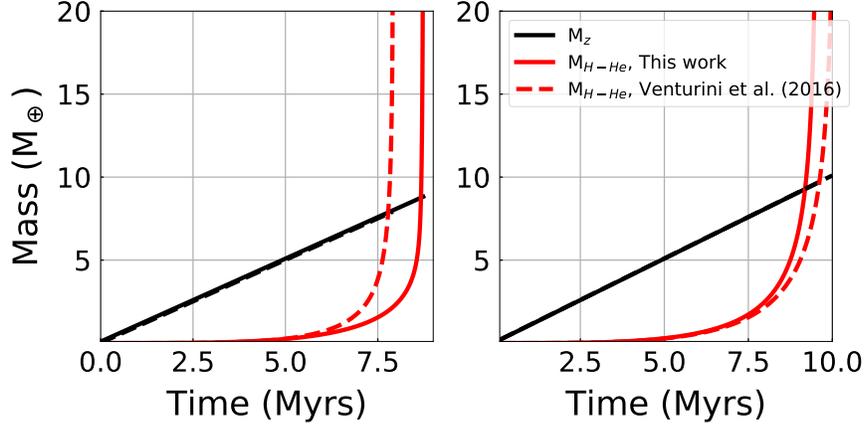}
	\caption{Growth of the planet as a function of time. The solid line represents our work, while the dashed line indicated the result from \citealt{Venturini16}. In the right panel plot is assumed an ideal equation of state.}
	\label{JuliaMesaComparison}
\end{figure}

We compare our code, which solves all the structure equations with the work of  \citealt{Venturini16} where the term \ref{dldr} is neglected. 
For this comparison, we assume a constant solid accretion rate. The results are shown in Figure~\ref{JuliaMesaComparison}. 
As can be seen from the left panel there is a good agreement between the two codes regarding the growth of the planet. 
In the right panel of Figure~\ref{JuliaMesaComparison} we perform the same simulation with using an ideal-gas equation of state and a constant opacity of $0.01$ $g/cm^2$ in order to ensure that the only difference between the two cases lies in the treatment of luminosity. The agreement between our simulations and this of \cite{Venturini16} provides an additional benchmark to our code.

\begin{figure}[h!]

	\centering
	\includegraphics[width=0.35\linewidth]{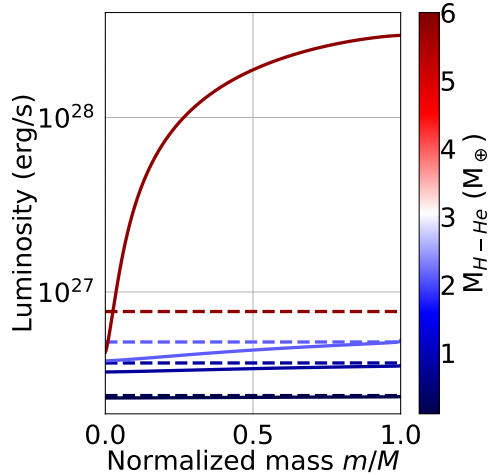}
	\caption{The planetary luminosity profile at four different times. Solid  lines indicate our work, while dashed ones are from \cite{Venturini16} assuming a constant luminosity. The total mass of the envelope is represented by a color scheme.}
	\label{lumonosityratio}
\end{figure}

Finally, in Figure~\ref{lumonosityratio} we show the luminosity profile within the planet's envelope at four different times. The dashed lines represent results from \cite{Venturini16}, while the solid lines indicate the luminosity profile derived with our code where the luminosity is not constant.  
It is clear from the figure that the constant luminosity approximation is inapplicable when the envelope's mass reaches $\sim 2-3$ M$_\oplus$.
We therefore suggest that in order to properly model the formation and evolution of gaseous-rich planets assuming a constant luminosity should be avoided.

\section{Attached phase - gas accretion rate.}
\label{Appendix -AttachedPhase}

\begin{figure}[h]

	\centering
	\includegraphics[width=0.75\linewidth]{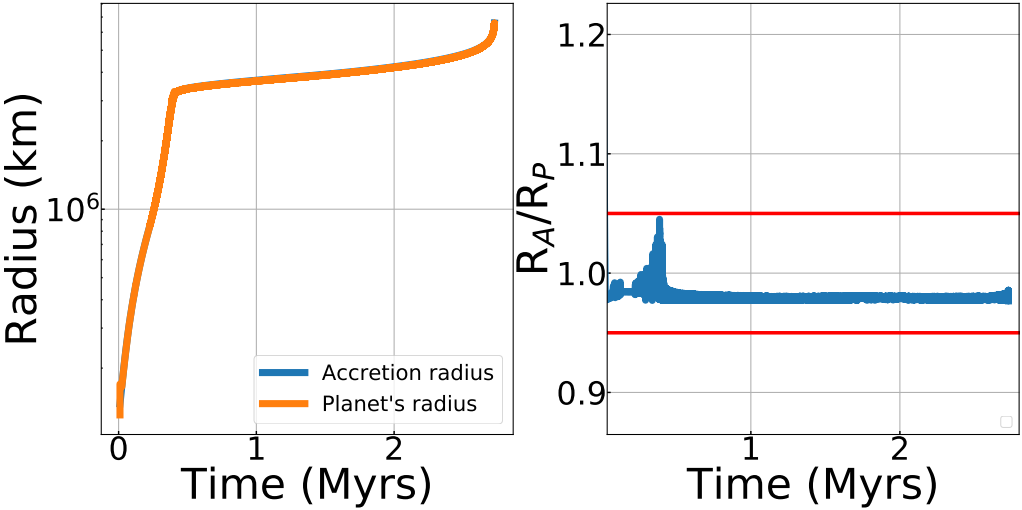}
	\caption{\textbf{Left:} Accretion radius (Eq~\ref{radius}) and planet's radius as a function of time. \textbf{Left:} Ratio of the accretion radius to the planet's radius as a function of time. The red lines indicate the 5\% tolerance.}
	\label{radiuscomparison}
\end{figure}

When the proto-planet mass is low ($\lesssim 30$ M$_\oplus$)  the gas accretion rate $\dot{M}_{xy}$ is obtained at each time-step through the requirement that the computed planet's radius $R_p$ equals to the accretion radius $R_A$ as shown in Equation~\ref{radius}.
This is done iteratively, adding different amounts of gas until convergence is reached where and $R_P$ equals $R_A$ within a given precision.
The gas accretion rate in this phase is determined using  the \verb!extras_check_model! subroutine.  
At the beginning of the time-step the heavy-element accretion rate is computed and the mass of the core is updated accordingly. 
At this point the radius is smaller than the one at the beginning of the time-step, we call it $R_{inter}$, representing an intermediate radius.
The first guess for the gas accretion rate is done is the following:
\begin{verbatim}
s % mass_change =out_rho *4./3.*pi*(-R_inter**3+R_acc**3)/(Msun*s%dt/secyer),
\end{verbatim}
where \verb!out_rho! is the density of the outermost layer.
At this point MESA solves the structure equations with the updated mass and composition of the  envelope. We then check whether the radius matches (within the precision given)  the accretion radius. 
If the two radii match, we proceed to the next time-step, otherwise the calculation is repeated assuming a different amount of gas, which depends whether the radius is smaller or larger than the accretion radius.
The \verb!extras_check_model! sub-routine is given by:
\begin{verbatim}
    if(abs(R_acc/R_inter -1) .ge.tollerance)then
        if(R_acc/R_inter .ge. 0)then 
             s% mass_change = s% mass_change *2
            extras_check_model = retry
         else 
            s% mass_change = s% mass_change /1.5
            extras_check_model = retry
    else 
        extras_check_model = keep_going
\end{verbatim}
The left panel of Figure~\ref{radiuscomparison} shows the planet's radius and the accretion radius (Equation~\ref{radius}) as a function of time. In this simulation, heavy-element enrichment is neglected and all the planetesimals are assumed to join the core. We use the solid accretion rate from \cite{pollack1996}, where the initial solid surface density is 10 g cm$^{-2}$ and the protoplanet is located at 5.2 AU. The tolerance for satisfying $R_A = R_P$ is set to  5\%. 
In the right panel of Figure~\ref{radiuscomparison} we show the ratio $R_A/R_P$, i.e., the accretion radius (Equation~\ref{radius}) divided by the computed planet's radius. The red lines indicate the limit of 5\%. As can be seen from the figure, the ratio is always within the used tolerance.

\begin{figure}[h]
	\centering
	\includegraphics[width=0.4\linewidth]{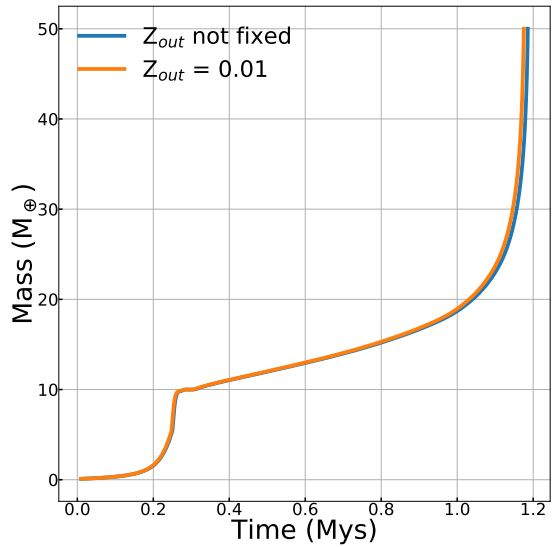}
	\caption{Mass of the planet as a function of time (including heavy-element enrichment). The blue line represents the case where we do not impose any constraint on the metallicity of the outermost layer, while the orange line indicate the case where the outermost layer is fixed to be $Z_{out} = 0.01$.}
	\label{outermostmetallicity}
\end{figure}

In principle, a change of the metallicity of the outermost layer could result in an artificial inflation of the radius, which in turn can affect the gas accretion rate. 
In Figure~\ref{outermostmetallicity} we show that imposing a fixed metallicity to the outermost layer does not significantly affect our result. 
The orange line represents the case where the metallicity of the outermost layer is fixed to $Z=0.01$, while the blue line indicate the case where no fixed metallicity is imposed on the outermost layer. We therefore conclude that our results for the planetary growth are robust.

\section{Detached phase - outer boundary conditions.}
\label{Appendix -DetatchedPhase}
During the first stage of the core accretion model, the total planetary mass is small ($ \le 50 $ M$_\oplus$) and it is attached to the disk. 
The outer envelope's temperature and pressure are fixed (therefore also the density) and are equal to the disk's temperature and pressure at the planet's location. 
When the computed gas accretion rate exceeds the limit that can be supplied by the disk, the planet detaches from the disk, and the gas accretion rate is significantly higher and is obtained via hydrodynamic formulae. The outer boundary conditions used in this case is the \verb!simple_photosphere! MESA option. The surface temperature, pressure, and density are obtained by solving the stellar structure equations using the Eddington $T(\tau)$ relation. 
Therefore, in this phase, the planetary surface temperature, pressure, and density vary with time. In addition, their actual values depend on the runaway gas accretion formula that is used And also whether the accretion energy can be deposited or is radiated away \citep{Berardo17,Cumming18}. 

We implemented three different prescriptions for the H-He accretion rate during  runaway as listed in Table~\ref{table2}. 
We fit with an high order polynomials the $\alpha = 4 \times 10^{-4}$ case of \cite{lissauer2009}. The fit is given by:
\begin{equation}
\label{Eq.fitexpression}
    \log\bigg(\frac{\dot{M}_{xy}}{\Sigma_g r_p^2 / P} \bigg) = \sum_{i=0}^{6} c_i \log^i \bigg(\frac{M_p}{M_\star}\bigg),
\end{equation}
where the coefficients are: $C_0 = 259.26$ , $C_1 = 584.32$, $C_2 = 478.51$, $C_3 = 195.81$, $C_4 = 43.25$, $C_5 = 4.94$, $C_6 = 0.229$.

\begin{figure}[h]
	\centering
	\includegraphics[width=0.75\linewidth]{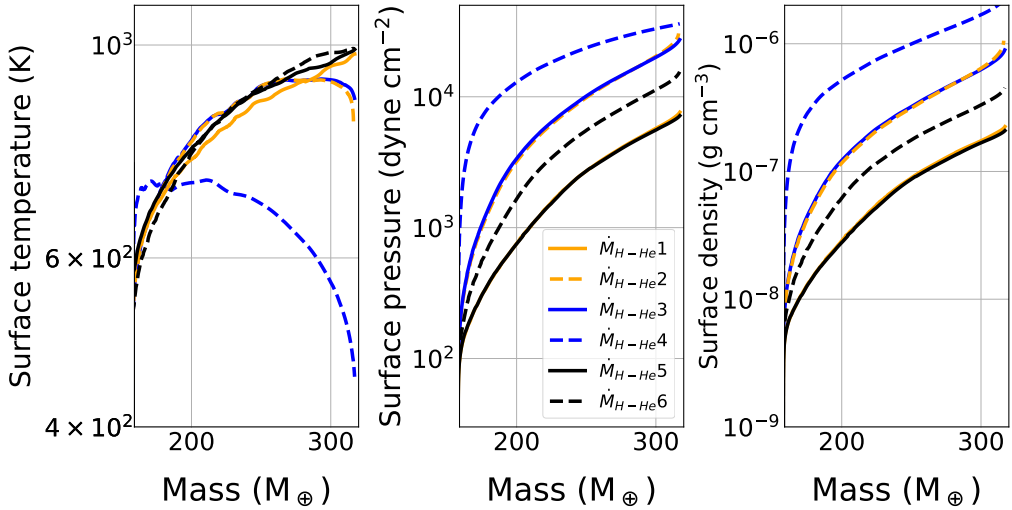}
	\caption{\textbf{Left:} Outermost temperature of the planet as a function of total planetary mass. The blue, orange and black line represents the solution of \cite{Ginzburg2019}, \cite{lissauer2009} and \cite{Kanagawa17}, respectively. Solid lines indicates a disk's viscosity of $4 \times 10^{-3}$, while the dashed one represents simulation in which the disk's viscosity is assumed to be $4 \times 10^{-4}$.
	\textbf{Middle:} Surface pressure of the planet as a function of total planetary mass.
	\textbf{Right:} Surface density of the planet as a function of total planetary mass.}
	\label{Fig.9-RunawayThermodynamics}
\end{figure}

The left panel of Figure~\ref{Fig.9-RunawayThermodynamics} shows the planet's surface temperature during the runaway accretion phase when assuming different formulae for the gas accretion rate as indicated by the different colors.  
The changes in the surface temperature are due to two reasons. On one hand, the gravitational energy provided by the accreted gas increases the surface temperature. On the other hand, the cooling of planet result in a loss of energy and therefore a decrease in its surface temperature. 
Therefore, the actual planetary temperature depend on the efficiency of these two processes.

It is clear that the planet's surface temperature does not remain constant also if the planet is not fully convective. This is different from the hypothesis in \cite{Ginzburg2019}, who assumed that the temperature changes from the initial value only when the planet's envelope becomes fully convective. 
The middle and right panels of the plot show the pressure and density of the planet, respectively.

\section{Dependence on material strength}
\label{section:materialstrength}

In this section, we investigate the sensitivity of our results on the assumed  material strength and latent heat of vaporization of the accreted planetesimals. 
We assume the material strengths and latent heat of vaporization for an icy (water) planetesimal to be equal to $10^6$ dyne $ $cm$^{-2}$ and $2.2 \times 10^{10}$ erg g$^{-1}$, respectively. A rocky planetesimal has a higher material  strength, which we set to $10^8$ dyne $ $cm$^{-2}$ and latent heat of vaporization equal to $8 \times 10^{10}$ erg g$^{-1}$. 
However, in both cases the heavy elements are represented by water for the EOS calculation.

\begin{figure}[h]
	\centering
	\includegraphics[width=0.7\linewidth]{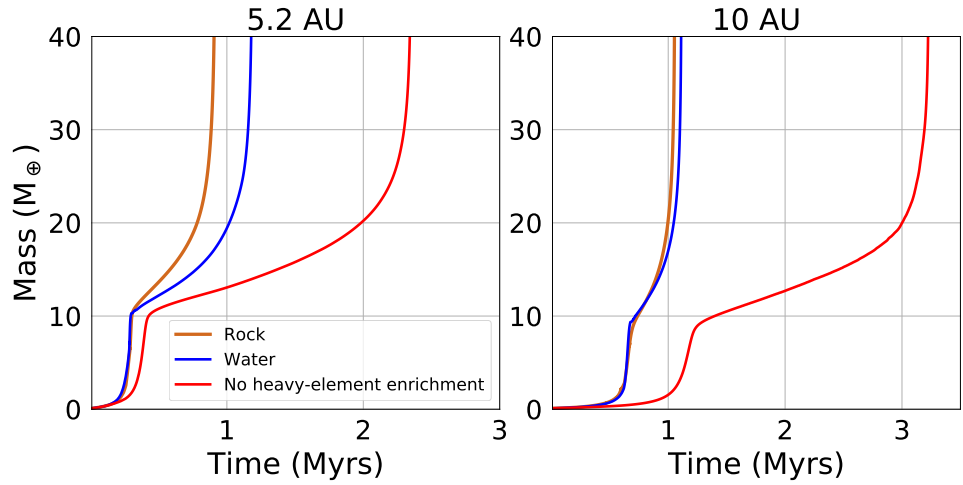}
	\caption{Planet's mass as a function of time. Blue and brown lines indicate a planetesimal made of water and rock, respectively. Red lines represent the case where all the solids are assumed to hit the core. The left and right panels correspond to formation locations of 5.2 AU and 10 AU, respectively.}
	\label{Figure15-parameters}
\end{figure}

Figure~\ref{Figure15-parameters} shows the planetary growth at two different planet's location assuming the \cite{pollack96} solid accretion rate. The blue and brown lines correspond to the planetesimal's composition of water and rock, respectively. The red lines represent the case where all the solids are assumed to reach the core.
We find that the growth history is rather similar for the two assumed planetesimals' compositions, where the planet's growth is slightly faster assuming rocky planetesimals. 
As expected, neglecting heavy-element enrichment results in a larger formation time-scale. Future studies should perform a detailed comparison when the different assumed heavy elements are also considered for the EOS and opacity calculation and we hope to address this in future work.  

\clearpage
\newpage
\bibliography{bibliography}

\end{document}